\documentclass[%
 reprint,
 amsmath,amssymb,
 aps,
]{revtex4-2}

\usepackage{graphicx}
\usepackage{dcolumn}
\usepackage{bm}

\begin{document}
\title{Does the high-energy AMS-02 positron flux originate from the dark matter density spikes around nearby black holes?}
\author{Man Ho Chan$^{\dag}$, Chak Man Lee}
\address{Department of Science and Environmental Studies, The Education University of Hong Kong, Tai Po, New Territories, Hong Kong, China
\\
Correspondence Email: $^{\dag}$chanmh@eduhk.hk\\
}
\begin{abstract}
Recent measurements made by the Alpha Magnetic Spectrometer (AMS) have detected accurate positron flux for energy range 1-1000 GeV. The energy spectrum can be best described by two source terms: the low-energy background diffusion term and an unknown high-energy source term. In this article, we discuss the possibility of the emission of positrons originating from dark matter annihilation in two nearby black hole X-ray binaries A0620-00 and XTE J1118+480. We show that the dark matter density spikes around these two black holes can best produce the observed AMS-02 high-energy positron flux due to dark matter annihilation with rest mass $m_{\rm DM} \approx 8000$ GeV via the $W^+W^-$ annihilation channel. This initiates a new proposal to account for the unknown high-energy source term in the AMS-02 positron spectrum.
\end{abstract}
 \maketitle
\section{Introduction}
The nature of dark matter remains one of the biggest mysteries in cosmology and particle physics. Although there are some studies claiming that dark matter with mass $m \sim 10-100$ GeV can account for the excess of gamma ray at the Galactic Center \cite{Daylan,Calore,Abazajian}, the excess of anti-proton in our Milky Way \cite{Cholis}, and the radio spectral signatures in other galaxies and galaxy clusters \cite{Chan,Chan2,Chan3}, these results are still inconclusive because the baryonic effects such as pulsar emissions are uncertain \cite{Malyshev}. Moreover, recent studies have placed stringent constraints on dark matter with mass $\sim 10-100$ GeV so that the available parameter space of the annihilation cross section \cite{Ackermann,Albert,Chan4} is getting much narrower.

Conversely, the cosmic-ray measurements by Alpha Magnetic Spectrometer (AMS) has revealed a certain excess of cosmic-ray flux at energy $1-1000$ GeV \cite{Aguilar,Aguilar2}. For the latest AMS-02 data release, the positron flux seems to be contributed by two source terms: a background cosmic-ray source and an unknown high-energy source \cite{Aguilar2}. Although invoking an ad hoc empirical form of pulsar emission \cite{Aguilar2} or some specific pulsar emission models \cite{Cholis2,Cholis3} can provide good agreements with the AMS-02 energy spectrum, other origins accounting for the unknown high-energy source are still possible. Some other studies have considered the dark matter annihilation model and primordial black hole model to explain the unknown source term \cite{Ghosh,Krommydas,Su}. However, the available parameter space of dark matter and primordial black hole evaporation has been severely constrained by other studies such as gamma-ray and cosmic-ray constraints \cite{Ackermann,Albert,Boudaud}.

Recently, it has been shown that two nearby black hole low mass X-ray binaries (BH-LMXBs), A0620-00 and XTE J1118+480, might contain dark matter density spikes \cite{Chan5,Ireland}. The high density of dark matter provides dynamical friction to slow down the companion stars which can satisfactorily explain the orbital period decay rate detected in A0620-00 and XTE J1118+480 \cite{Chan5}. If the black holes in these binaries are primordial in nature, dark matter density spikes could be formed around these primordial black holes \cite{Mack,Ricotti,Ireland}. If dark matter in these binaries can self-annihilate to give out high-energy particles (e.g. electrons, positrons, neutrinos), these particles could diffuse out and contribute to the background cosmic rays. We expect that some of the positrons detected by AMS might originate from dark matter annihilation in nearby black hole systems. Some previous studies have also explored the possibility of constraining dark matter using radio data of these two binaries \cite{Kar}. In this article, we explore the possibility whether dark matter annihilation in A0620-00 and XTE J1118+480 could explain the unknown source term revealed in the AMS-02 spectral data.

\section{Cosmic rays contributed by dark matter annihilation}
In the followings, we present the theoretical framework for modelling the cosmic rays contributed by dark matter annihilation. We assume that most of the dark matter contributed cosmic-ray flux detected by AMS originates from the nearby black hole binaries A0620-00 and XTE J1118+480 (i.e. neglecting the contribution due to galactic dark matter). These binaries are located at $1.06\pm 0.12$ kpc and $1.70 \pm 0.10$ kpc from us respectively \cite{Gonzalez}. The dark matter density profiles surrounding the corresponding black holes can be described by the following spike profile \cite{Chan5,Lacroix}:
\begin{equation}
\rho_{\rm DM}=\left\{
\begin{array}{ll}
0 & {\rm for }\,\,\, r\le 2R_s \\
\rho_0 \left(\frac{r}{r_{\rm sp}}\right)^{-\gamma_{\rm sp}} & {\rm for }\,\,\, 2R_s <r \le r_{\rm sp}, \\
\rho_0 & {\rm for}\,\,\, r> r_{\rm sp} \\
\end{array}
\right.
\end{equation}
where $R_s=2GM_{\rm BH}/c^2$ with $M_{\rm BH}$ being the black hole mass. The empirical parameters $r_{\rm sp}$, $\rho_0$ and $\gamma_{\rm sp}$ for each black hole binary have been obtained in \cite{Chan5}.

If dark matter particles would self-annihilate to give high-energy electrons and positrons, the energy spectrum of the electrons or positrons produced is given by \cite{Vollmann}
\begin{equation}
Q(E,r)=\frac{\langle \sigma v \rangle \rho_{\rm DM}^2}{2m_{\rm DM}^2}\frac{dN_{\rm e,inj}}{dE},
\end{equation}
where $m_{\rm DM}$ is the mass of a dark matter particle, $\langle \sigma v \rangle$ is the annihilation cross section and $dN_{\rm e,inj}/dE$ is the injected energy spectrum of the dark matter annihilation. Different injected energy spectra can be calculated numerically for different annihilation channels \cite{Cirelli}. The electrons and positrons produced from dark matter annihilation would diffuse out from the black hole dark matter density spike regions and cool down mainly due to synchrotron emission and inverse Compton scattering (ICS). The diffusion and cooling process can be modeled by the following diffusion-cooling equation:
\begin{eqnarray}
\frac{\partial}{\partial t}\frac{dn_{\rm e}}{dE}&=&\frac{D(E)}{r^2}\frac{\partial}{\partial r}\left(r^2\frac{\partial }{\partial r}\frac{dn_{\rm e}}{dE}\right)+\frac{\partial}{\partial E}\left[b_{\rm T}(E)\frac{dn_{\rm e}}{dE}\right]\nonumber\\
&&+Q(E,r),
\label{diffusion}
\end{eqnarray}
where $dn_{\rm e}/dE$ is the energy spectrum of the electrons or positrons after cooling and diffusion, $b(E)$ is the total cooling rate, and $D(E)=D_0(E/\rm 1~GeV)^{\delta}$ is the diffusion function. For the Milky Way, recent studies show that the best diffusion coefficient is $D_0= 4.0^{+0.6}_{-0.5}\times 10^{28}$ cm$^2$/s \cite{Wu}. Also, the best diffusion index $\delta=0.31 \pm 0.04$ constrained \cite{Wu} is very close to the prediction of the benchmark Kolmogorov model $\delta=1/3$ \cite{Kolmogorov} and agrees with previous calculations \cite{Korsmeier}. Here, we adopt $D_0=4 \times 10^{28}$ cm$^2$/s and $\delta=1/3$.

For the cooling function $b(E)$, the contributions due to synchrotron emission and ICS are given by \cite{Siffert}
\begin{equation}
b(E)=0.0254E^2B^2 + 1.02U_{\rm rad}E^2,
\label{cooling}
\end{equation}
where $B$ in the unit of $\mu$G is the magnetic field strength and $U_{\rm rad}$ in the unit of eV/cm$^3$ is the radiation density. Previous studies have shown that the magnetic field strength near the position of our solar system is $\approx 2$ $\mu$G \cite{Sun,Beck}. Since the binaries are nearby us, we approximate a constant magnetic field strength 2 $\mu$G throughout the diffusion of the electrons and positrons from the binaries to us. For the radiation density, we take a superposition of three blackbody-like spectra for cosmic microwave background, infrared light and visible light \cite{Cirelli2}:
\begin{equation}
U_{\rm rad}=\int_0^{\infty} \sum_{i=1}^3 \mathcal{N}_i \frac{8\pi\epsilon^3}{h^3c^3}\frac{d\epsilon}{\exp(\epsilon/kT_i)-1},
\end{equation}
with the parameters for the disk region used in \cite{Cirelli2}.

By setting the boundary conditions $(\partial/\partial t)(dn_e/dE)=0$ and $dn_e(r_h,E)/dE=0$ with the diffusion halo radius $r_h$, the general solution of the equilibrium electron density spectrum can be obtained in terms of the Fourier-series representation of the Green's function as follows \cite{Vollmann}:
\begin{eqnarray}
\frac{dn_e}{dE}&=&\sum_{n=1}^{\infty}\frac{2}{b_T(E)r_h}\frac{\sin\left(\frac{n\pi r}{r_h}\right)}{r}
\nonumber\\
&&\times\int_E^{^{m_{\rm DM}}}dE'e^{-n^2[\eta(E)-\eta(E')]}\nonumber\\
&&\times\int_0^{r_h}dr'r'\sin\left(\frac{n\pi r'}{r_h}\right)Q(E',r'),
\label{dndE}
\end{eqnarray}
where $r$ denotes the distance to the black hole binaries and the dimensionless variable $\eta(E)$ is given by
\begin{eqnarray}
\eta(E)&=&\frac{1}{1-\delta}\left(\frac{6.42\pi\;{\rm kpc}}{r_h}\right)^2\left(\frac{D_0}{10^{28}{\rm cm^2/s}}\right)
\nonumber\\
&&\times\left(\frac{1}{1+\left({B}/{3.135 \;\mu{\rm G}}\right)^2}\right)\left(\frac{1\;{\rm GeV}}{E}\right)^{1-\delta}.
\end{eqnarray}

The cosmic-ray flux of the electrons or positrons (in cm$^{-2}$ s$^{-1}$ sr$^{-1}$) produced by dark matter annihilation from the black hole binaries is thus given by
\begin{equation}
\Phi_{\rm DM}=\frac{c}{4\pi} \frac{dn_e}{dE}.
\end{equation}
In this analysis, we assume that the electrons and positrons produced from dark matter annihilation only originate from the two nearby black hole binaries A0620-00 and XTE J1118+480. Generally speaking, there are three free parameters ($r_h$, $m_{\rm DM}$, $\langle \sigma v \rangle$) involved in the dark matter annihilation model. Nevertheless, both parameters $r_h$ and $\langle \sigma v \rangle$ are proportional to the cosmic-ray flux so they are degenerate with each other. Therefore, we choose the benchmark value $r_h=20$ kpc \cite{Ding} to reduce the number of free parameters to two (i.e. $m_{\rm DM}$ and $\langle \sigma v \rangle$) in our analysis. Therefore, the value of $\langle \sigma v \rangle$ here represents the effective annihilation cross section only, but not the actual annihilation cross section for dark matter.

\section{Data analysis}
In the followings, we will examine the AMS-02 spectral data by considering the dark matter contribution and the background model. As discussed in \cite{Aguilar2}, the low-energy part of the positron flux, mainly at energy $E \sim 1-100$ GeV, should originate from the positrons produced in the collisions of ordinary cosmic rays with the interstellar gas. This background diffuse term can be modeled by \cite{Aguilar2,Cavasonza}
\begin{equation}
\Phi_{\rm d}=C_{\rm d}\frac{E^2}{\tilde{E}^2}\left(\frac{\tilde{E}}{E_1}\right)^{\gamma_{\rm d}}
\label{bgdwithDM}
\end{equation}
where $\tilde{E}=E+\phi$, and the constant $E_1$ is chosen to be 7.0 GeV. Here, $C_{\rm d}$, $\phi$, and $\gamma_{\rm d}$ are free parameters for model fitting.

Adding the background diffuse term with the dark matter contribution, we get the total positron flux: $\Phi_{e^+}=\Phi_{\rm d}+\Phi_{\rm DM}$. There are five free parameters involved in the model: $C_{\rm d}$, $\phi$, $\gamma_{\rm d}$, $m_{\rm DM}$ and $\langle \sigma v \rangle$. To get the best-fit parameters, we minimize the reduced $\chi^2$ value, which is defined as
\begin{equation}
\chi_{\rm red}^2=\frac{1}{N-5} \sum_{i=1}^{N} \frac{(\Phi_{e^+}-\Phi_i)^2}{\sigma_i^2},
\end{equation}
where $\Phi_i$ and $\sigma_i$ are the observed positron flux and its uncertainty respectively.

By fitting the AMS positron spectrum with our model using different annihilation channels, we find that there are two channels giving relatively good fits: $W^+W^-$ and $\tau^+\tau^-$. In Fig.~1, we show the positron spectra for these two channels using the best-fit parameters. The $\chi_{\rm red}^2$ value for the $W^+W^-$ and $\tau^+\tau^-$ channels are 0.64 and 1.51 respectively, which suggests that the fits are very good, especially for the $W^+W^-$ channel. For different models suggesting pulsar emission to account for the source term, the $\chi_{\rm red}^2$ value for the best model is $\approx 1.0$ \cite{Cholis2}. Therefore, our model assuming dark matter annihilating via the $W^+W^-$ channel can provide a smaller $\chi_{\rm red}^2$ value and thus offer an alternative possible account for the AMS-02 positron spectrum.

We also test whether our model would affect the DAMPE spectrum (electrons and positrons) as well \cite{Ambrosi}. By considering the dark matter emission with the DAMPE background model in \cite{Ambrosi,Liu}, we can fit with the DAMPE spectrum (see Fig.~2). By modifying the free parameters in the DAMPE background model, adding dark matter emission can give $\chi_{\rm red}^2=0.42$ (the outlying data point at 1.4 TeV has not been included), which suggests that the fit is very good. In other words, our model generally agrees with both AMS-02 and DAMPE spectra.

Moreover, we have also checked with the isotropic gamma-ray constraints. For our model, the maximum gamma-ray flux due to dark matter annihilation is $\sim 10^{-6}$ GeV m$^{-2}$ s$^{-1}$ sr$^{-1}$, which is smaller than the observed background isotropic gamma-ray flux $\sim 10^{-5}$ GeV m$^{-2}$ s$^{-1}$ sr$^{-1}$ measured by Fermi-LAT \cite{Ackermann2} (see Fig.~3). This means that our model does not violate current stringent isotropic gamma-ray constraint.

\section{Discussion}
In this article, by using the fact that two nearby black hole X-ray binaries (A0620-00 and XTE J1118+480) contain dark matter density spikes, the positron emission due to dark matter annihilation can provide a very good fit to the AMS-02 positron spectrum. The best resultant fit (assuming dark matter annihilating via the $W^+W^-$ channel) is better than that using different pulsar emission models. Also, adding the dark matter emission component can still provide a good fit to the DAMPE spectrum and does not violate the background isotropic gamma-ray constraint. This suggests that our model can best explain the unknown high-energy source term appeared in the AMS-02 positron spectrum. Therefore, the high-energy positrons detected by the AMS likely originate from the two nearby black hole X-ray binaries via dark matter annihilation in these systems.

Based on our results, the best-fit dark matter mass is $m_{\rm DM}\approx 8000$ GeV, which is heavier than our normal expectation. This agrees with some of the recent suggestions that the dark matter mass is larger than 1000 GeV \cite{Chan4,Calore2}. This provides a new hint for searching new particles in the Large Hadron Collider experiments. Apart from positron detection, considering other kinds of detection such as neutrinos, gamma rays, and radio spectra can provide a much more stringent constraint on the dark matter mass. Therefore, future multi-messenger studies would help reveal the nature of dark matter and the specific properties of the interaction between dark matter and nearby black holes.

\section{Data Availability Statement}
The data the support the findings of this article are openly available \cite{Aguilar2,Ambrosi,Ackermann2}.

\begin{acknowledgments}
The work described in this paper was partially supported by the Dean's Research Fund and the grant from the Research Grants Council of the Hong Kong Special Administrative Region, China (Project No. EdUHK 18300922 and EdUHK 18300324).
\end{acknowledgments}

\begin{figure}
\vskip 3mm
\includegraphics[width=85mm]{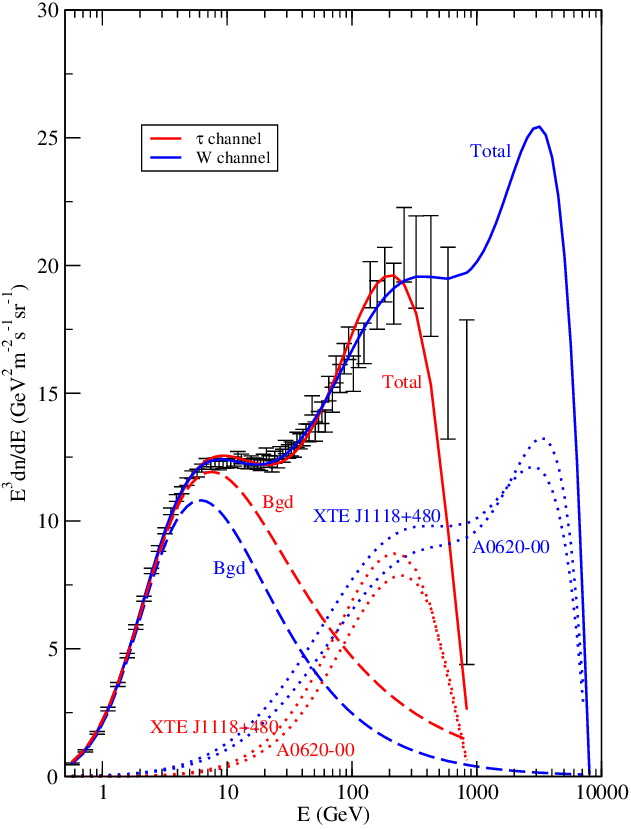}
\caption{Fitting the AMS-02 positron spectrum with the background emission model plus dark matter annihilation model. The blue and red solid lines represent the best fits for the $W^+W^-$ channel with $m_{\rm DM}=8000$ GeV and $\tau^+\tau^-$ channel with $m_{\rm DM} = 900$ GeV respectively. The dashed lines represent the background emission components and the dotted lines represent the corresponding components originating from A0620-00 and XTE J1118+480 binaries. The data of the AMS-02 positron spectrum are extracted from \cite{Aguilar2}.}
\label{Fig1}
\vskip 3mm
\end{figure}

\begin{figure}
\vskip 3mm
\includegraphics[width=85mm]{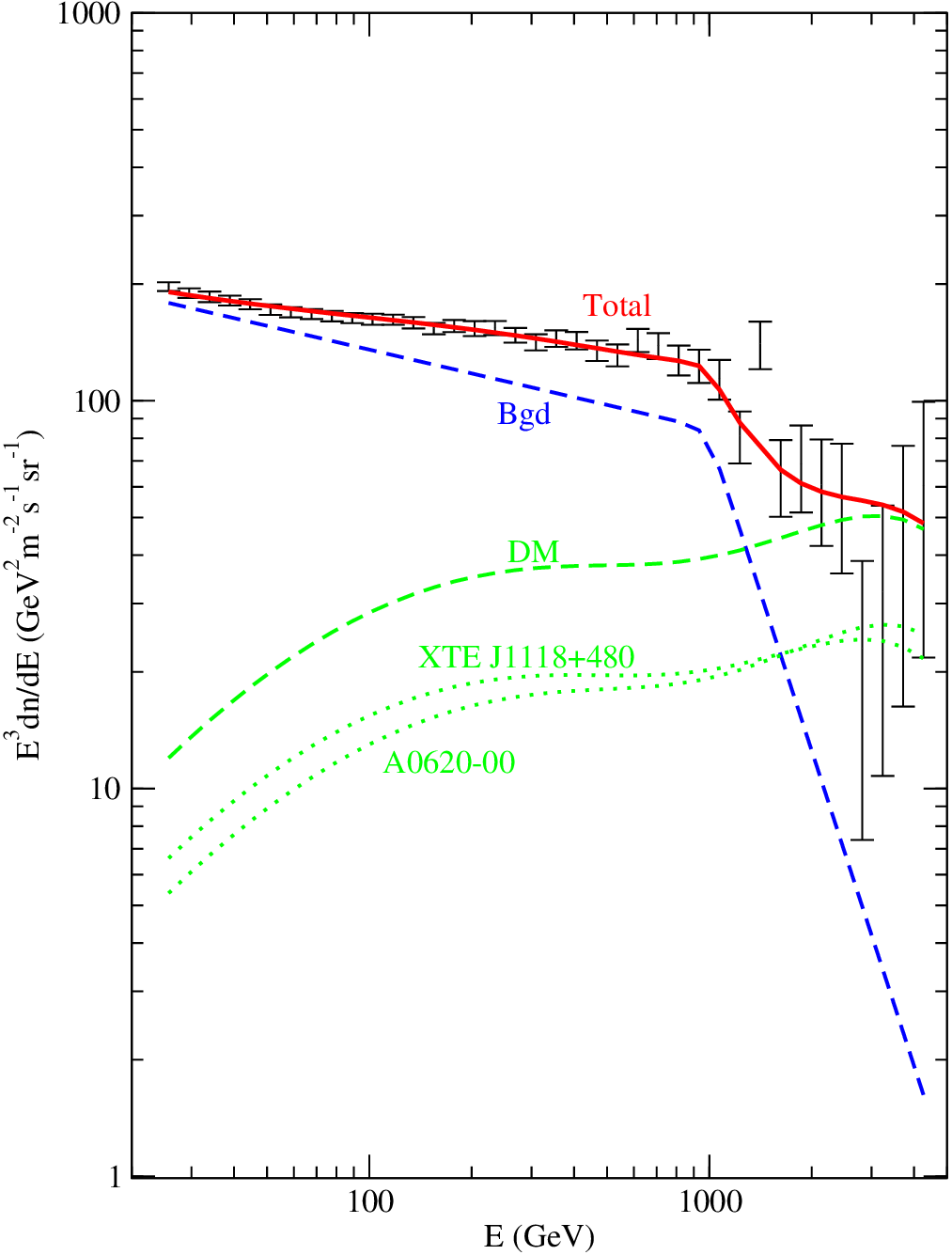}
\caption{Fitting the DAMPE spectrum (electron and positron flux) with the background emission model plus dark matter annihilation model. The red solid line represents the total electron and positron flux. The blue and green dashed lines represent the background emission component and the dark matter ($m_{\rm DM}=8000$ GeV) component respectively. The green dotted lines indicate the individual components originating from A0620-00 and XTE J1118+480 binaries. The data are extracted from \cite{Ambrosi}.}
\label{Fig2}
\vskip 3mm
\end{figure}

\begin{figure}
\vskip 3mm
\includegraphics[width=85mm]{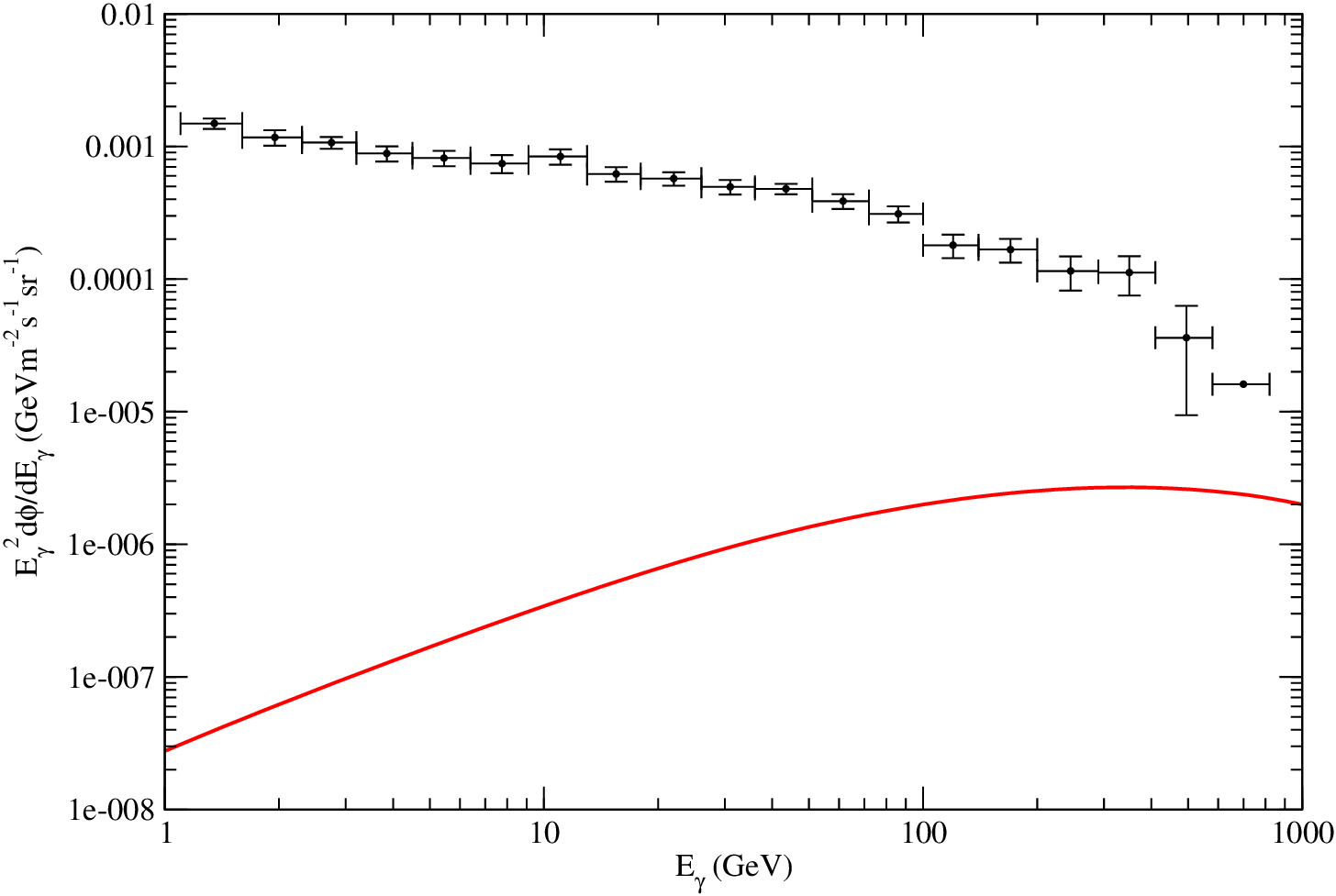}
\caption{The red line represents the gamma-ray flux of the dark matter annihilation model following the best-fit scenario ($W^+W^-$ channel with $m_{\rm DM}=8000$ GeV). The error bars indicate the isotropic gamma-ray flux from the Fermi-LAT data extracted from Table 3 of \cite{Ackermann2}.
}
\label{Fig3}
\vskip 3mm
\end{figure}

\end{document}